\documentclass[%
 reprint,
 amsmath,amssymb,
 aps,
]{revtex4-2}

\usepackage{graphicx}
\usepackage{dcolumn}
\usepackage{bm}

\begin{document}

\preprint{APS/123-QED}

\title{Mn intercalation induced a ferrimagnetic to ferromagnetic \\ transition in trigonal Cr$_{5}$Te$_{8}$ single crystal}

\author{Ze-Xin Liu, Yu Liu, Sen-Miao Zhao,  De-Wei Zhao, Li Ma, Deng-Lu Hou}

\author{Guo-Ke Li}%
 \email{Corresponding author: liguoke@126.com}
\affiliation{%
Hebei Advanced Thin Films Laboratory, College of Physics, Hebei Normal University, Shijiazhuang, 050024, China.
}%

\date{\today}

\begin{abstract}
Tailoring the magnetic properties of chromium tellurides via heterointercalation with extrinsic transition metals remains largely unexplored. Here, we report a comprehensive investigation of trigonal Cr$_5$Te$_8$ and Cr$_4$MnTe$_8$ single crystals, in which Mn substitution elevates the magnetic ordering temperature from 226 to 249 K and enhances the saturation magnetic moment per magnetic ion ($m_{\text{S}}$) from 2.00 to 2.66 $\mu_{\text{B}}$ at 5 K. Remarkably, the observed $m_{\text{S}}$ enhancement significantly exceeds the contribution of Mn ion moments alone, indicating the relief of intrinsic spin compensation within the parent lattice. First-principles calculations definitively establish that pristine $\text{Cr}_5\text{Te}_8$ exhibits ferrimagnetic ordering with a computed $m_\text{S}$ of 1.98~$\mu_\text{B}$, and further reveal that preferential occupation of the van der Waals gaps by Mn ions induces a ferrimagnetic-to-ferromagnetic transition, yielding a predicted $m_\text{S}$ of 2.94~$\mu_\text{B}$. These findings not only resolve the magnetic ground state of trigonal Cr$_5$Te$_8$ but also identify heterointercalation as a robust strategy for engineering the spin textures of chromium tellurides.
\end{abstract}

\maketitle


\section{INTRODUCTION}

Since the successful exfoliation of 2D van der Waals (vdW) ferromagnets\cite{Gong2017,Huang2017,Deng2018}, chromium tellurides (Cr$_x$Te$_y$) have emerged as a compelling platform for exploring intrinsic room-temperature ferromagnetism and pronounced magnetic anisotropy in layered quantum materials\cite{Zhang2020,Zhang2021,Wang2024,Zhang2025,Huang2026}. A defining characteristic of this family is the self-intercalation of excess Cr atoms into the parent CrTe$_2$ vdW gaps, yielding the general formula Cr$_{1+\delta}$Te$_2$ ($0 \leq \delta \leq 1$) , where the intercalation parameter $\delta$ serves as a decisive ``tuning knob'' for modulating magnetic exchange couplings\cite{Zhang2020,Fujisawa2020,Liu2022,Yang2023,Fujisawa2023}. While heterointercalation has successfully engineered exotic quantum states in other metal dichalcogenides\cite{Parkin1980,Parkin1980b,Morosan2006,Rajapakse2021,Zhou2025}, such as inducing superconductivity in TiSe$_2$ \cite{Morosan2006}, or introducing ferromagnetism in MoTe$_2$\cite{Zhou2025}, its application in the Cr$_x$Te$_y$ system remains severely constrained. Prior efforts have predominantly focused on self-intercalation \cite{Zhang2020,Fujisawa2020,Liu2022,Yang2023,Fujisawa2023}, which is intrinsically constrained by the fixed Cr-Cr exchange coupling and unable to circumvent the inherent limitations in magnetic tunability imposed by the $\text{Cr}_x\text{Te}_y$ phase diagram\cite{Wang2019,Wang2024,Wang2024b,Lan2025}.

Trigonal Cr$_5$Te$_8$ (tr-Cr$_5$Te$_8$) represents an ideal, yet paradoxically complex, testbed for addressing this gap due to its well-defined vdW gaps and amenability to high-quality single-crystal growth\cite{Wang2019,Liu2018,Luo2018}. Despite its structural advantages, the magnetic ground state of tr-Cr$_5$Te$_8$ remains a subject of intense debate\cite{Mondal2019,Wang2019}. The experimental saturation moments ($1.57$--$2.25$ $\mu_{\text{B}}$/Cr) fall drastically short of theoretical ionic predictions ($\sim3.0$ $\mu_{\text{B}}$/Cr)\cite{Mondal2019,Wang2019,Tang2022,Jiang2022}, contrasting sharply with the moments observed in its monoclinic counterpart ($\sim2.4$ $\mu_{\text{B}}$/Cr)\cite{Jiang2022-a}. This discrepancy suggests either partial antiferromagnetic ordering or geometric frustration within the Cr sublattice, necessitating further experimental validation\cite{Tang2022,Jiang2022-a}. Although transition metal doping with Ti and V in polycrystalline samples has been attempted\cite{Hatakeyama2000,Hatakeyama2002-1,Hatakeyama2002-2,Huang2005}, these dopants typically quench magnetic order, and the inherent grain disorder precludes precise investigation of how foreign ions reconfigure the intrinsic exchange coupling and magnetocrystalline anisotropy. Motivated by the ability of Mn to mediate robust long-range spin correlations and the predicted strong Mn--Cr ferromagnetic coupling\cite{Li2005,Li2011,Debnath2016,Zhang2018,Maniv2020}, we propose Mn heterointercalation into tr-Cr$_5$Te$_8$ single crystals. Furthermore, the ionic radius of Mn$^{3+}$ ($\sim0.64$~\AA) is compatible with Cr$^{3+}$ ($\sim0.61$~\AA)\cite{Tiwari2019}, enabling an optimal balance between structural stability and magnetic interaction efficiency.

In this work, we present a comparative investigation of pristine and Mn-substituted tr-Cr$_5$Te$_8$ single crystals. Combining magnetometry with first-principles calculations, we establish that Mn ions selectively occupy vdW interstitial sites, functioning as a ``magnetic switch'' that elevates the magnetic ordering temperature ($T_\text{M}$) from 226 to 249 K and the saturation moment per magnetic ion from 2.00 to 2.66~$\mu_\text{B}$. This dramatic enhancement provides definitive evidence that tr-Cr$_5$Te$_8$ inherently hosts a ferrimagnetic ground state, which Mn intercalation reconfigures into a ferromagnetic order. By resolving the magnetic ground-state ambiguity of tr-Cr$_5$Te$_8$ and establishing heterointercalation as a protocol for magnetic engineering, our work provides a general paradigm to optimize the magnetic performance of $\text{Cr}_x\text{Te}_y$-based 2D vdW materials.

\section{Experimental}
Single crystals of pristine and Mn-substituted tr-Cr$_5$Te$_8$ were grown using the Te-flux method. High-purity elemental precursors with nominal molar ratios of Cr:Te = 15:85 and Cr:Mn:Te = 12:8:80 were weighed and sealed in evacuated quartz tubes. The sealed quartz tubes were heated to 1100~$^\circ$C, equilibrated for 24~h, and then cooled slowly to 650~$^\circ$C at a rate of 2~$^\circ$C/h. Excess Te flux was removed by centrifugation, yielding shiny, plate-like single crystals, as shown in the insets of Figs.~1(a) and (b). The chemical composition and crystal structure of the samples were characterized by energy-dispersive X-ray spectroscopy (EDS, XFlash7\textbar 60) and X-ray diffraction (XRD, PANalytical). Magnetic and electrical transport properties were measured using a physical property measurement system (MPMS-9, Quantum Design).

First-principles calculations were performed using the Vienna \textit{Ab initio} Simulation Package (VASP) with a plane-wave energy cutoff of 400~eV and a $\Gamma$-centered $5 \times 5 \times 3$ Monkhorst-Pack $k$-point mesh. A $1 \times 1 \times 2$ supercell was constructed based on experimental lattice parameters, which were held fixed during relaxation, while all atomic positions were fully relaxed to minimize the total energy. The magnetic ground state was identified by comparing the total energies of ferromagnetic and ferrimagnetic configurations with spin orientations varied from $0^\circ$ to $90^\circ$ relative to the $c$-axis. Spin-orbit coupling was incorporated in the self-consistent iterations, reaching convergence with a total energy tolerance of $10^{-6}$~eV and a force threshold of 0.1~meV/\AA.

\section{Results and discussion}
\begin{figure}[htbp!]
    \centering
    \includegraphics[width=0.5\textwidth]{Fig.1.jpg}
    \vspace{0pt} 
    \caption{ Compositional and structural characterization of pristine and Mn-intercalated Cr$_5$Te$_8$ single crystals. Rietveld-refined powder XRD patterns of (a) pristine Cr$_5$Te$_8$ and (b) Mn-substituted Cr$_5$Te$_8$ single crystals, both of which are indexed to the trigonal $P\bar{3}m1$ space group. The insets show optical micrographs, the corresponding EDS elemental maps, and the derived atomic percentages indicating the chemical compositions of Cr$_5$Te$_8$ and Cr$_4$MnTe$_8$, respectively. XRD scans acquired from the cleaved surfaces of (c) pristine Cr$_5$Te$_8$ and (d) Mn-substituted Cr$_5$Te$_8$ single crystals, exhibiting sharp (000$l$) reflections that confirm the quasi-2D nature of the crystals.}
    \vspace{0pt} 
    \label{Fig. 1}
\end{figure}

Rietveld refinement of the powder XRD patterns shown in Figs.~1(a) and (b) confirms that both the pristine and Mn-substituted crystals are phase-pure and crystallize in a trigonal symmetry with the $P\bar{3}m1$ space group. Upon Mn substitution, the lattice parameters exhibit a significant expansion, increasing from $a = b = 7.81$~\AA, $c = 11.98$~\AA\  to $a = b = 7.83$~\AA, $c = 12.17$~\AA\cite{Wang2024}. This evolution is primarily attributed to the substitution of the Cr$^{3+}$ ions ($\sim0.61$~\AA) by the larger Mn$^{3+}$ ions ($\sim0.63$~\AA) within the host lattice. Complementary EDS mapping presented in the insets of Figs.~1(a) and (b) further verifies the homogeneous elemental distribution and the absence of phase segregation in both samples. Quantitative analysis yields actual compositions of Cr$_{4.82}$Te$_{8.00}$ and Cr$_{4.11}$Mn$_{0.77}$Te$_{8.00}$, and these two samples are denoted as Cr$_5$Te$_8$ and Cr$_4$MnTe$_8$ hereafter to align with conventional notation and facilitate subsequent first-principles calculations. Finally, XRD scans collected from naturally cleaved crystal surfaces, depicted in Figs.~1(c) and (d), exclusively exhibit sharp (000$l$) reflections. This observation confirms that the $c$-axis is oriented perpendicular to the crystal facet, reflecting the preferential cleavage parallel to the $ab$-plane, a defining characteristic of quasi-2D vdW materials\cite{Huang2026}.

\begin{figure}[htbp!]
    \centering
    \includegraphics[width=0.5\textwidth]{Fig.2.jpg}
    \vspace{0pt} 
    \caption{Magnetic properties of trigonal Cr$_5$Te$_8$ and Cr$_4$MnTe$_8$ single crystals. (a, b) Temperature-dependent magnetization curves measured under zero-field-cooled (ZFC) and field-cooled (FC) protocols with a 100~Oe magnetic field applied parallel and perpendicular to the $c$-axis for Cr$_5$Te$_8$ and Cr$_4$MnTe$_8$, respectively. (c, d) Isothermal magnetization hysteresis loops recorded at 5~K with the magnetic field applied parallel and perpendicular to the $c$-axis for Cr$_5$Te$_8$ and Cr$_4$MnTe$_8$, respectively, which reveal pronounced magnetocrystalline anisotropy in both systems. The insets show the temperature dependence of the saturation magnetization extracted from the 90~kOe data.}
    \vspace{0pt} 
    \label{Fig. 2}
\end{figure}

Temperature-dependent magnetization measurements were performed on tr-Cr$_5$Te$_8$ single crystals under a 100~Oe magnetic field applied both parallel to the $c$-axis and within the $ab$-plane, as presented in Fig.~2(a). For the magnetic field along the $c$-axis, the sample exhibits a magnetic ordering temperature ($T_{\text{M}}$) of 226~K. The pronounced bifurcation between zero-field-cooled (ZFC) and field-cooled (FC) magnetization curves below 220~K signifies robust magnetocrystalline anisotropy or the presence of spin canting\cite{Wang2019,Mondal2019,Wang2024}. Conversely, for the magnetic field perpendicular to the $c$-axis, both the ZFC and FC curves nearly overlap and display a distinct cusp at approximately 238~K. This feature is characteristic of predominantly antiferromagnetic ordering within the $ab$-plane, indicating a complex coexistence of interlayer ferromagnetic and intralayer antiferromagnetic components in excellent agreement with previous studies on tr-Cr$_5$Te$_8$\cite {Luo2018}. In the case of Mn-substituted Cr$_4$MnTe$_8$ shown in Fig.~2(b), the ZFC and FC magnetization curves show nearly perfect overlap across both field orientations, with only minor splitting observed below 30~K. Critically, the $T_{\text{M}}$ is substantially enhanced by 23~K relative to the pristine sample, reaching 249~K. This elevation suggests that Mn ions reconfigure the exchange network, effectively suppressing the antiferromagnetic components within the $ab$-plane while strengthening the ferromagnetic coupling\cite{Li2005,Li2011}. Additionally, a pronounced anomaly emerges at approximately 225~K, where the in-plane magnetization exceeds the out-of-plane value. This crossover behavior strongly suggests the occurrence of a temperature-driven spin reorientation transition, a phenomenon wherein the magnetic easy axis flips from the $c$-axis toward the $ab$-plane upon warming\cite{Wang2024}.

Isothermal magnetization loops measured at 5~K for both crystals, shown in Figs.~2(c) and (d), reveal that the magnetization along the $c$-axis reaches saturation at markedly lower fields than that in the $ab$-plane. This confirms the $c$-axis as the magnetic easy axis, a behavior indicative of robust perpendicular magnetic anisotropy that is characteristic of quasi-2D Cr$_x$Te$_y$\cite{Wang2024b,Zhang2020}. Mn substitution induces pronounced modifications to the magnetic behavior, in which the coercivity along the $c$-axis at 5~K decreases from 269~Oe to 47~Oe, and the saturation magnetization rises from 9.64~$\mu_{\text{B}}$/f.u. to 12.98~$\mu_{\text{B}}$/f.u. This corresponds to a significant increase in the saturation magnetic moment per magnetic ion ($m_{\text{S}}$) from 2.00 to 2.66~$\mu_{\text{B}}$. Remarkably, the $m_{\text{S}}$ value of Cr$_4$MnTe$_8$ not only exceeds that reported for monoclinic Cr$_5$Te$_8$ ($\sim2.4$~$\mu_{\text{B}}$/Cr)\cite{Jiang2022-a}, but also exceeds the upper limit characteristic of the CrTe phase ($\sim2.5$~$\mu_{\text{B}}$/Cr)\cite{Ohta1993,Kanomata2000,Wu2021}. The observed enhancement of 0.66~$\mu_{\text{B}}$ per magnetic ion far surpasses the theoretical contribution anticipated from simple high-spin Mn$^{3+}$ ($d^4$, 4.0~$\mu_{\text{B}}$) or Mn$^{2+}$ ($d^5$, 5.0~$\mu_{\text{B}}$) substitution. This discrepancy strongly implies that pristine tr-Cr$_5$Te$_8$ hosts substantial spin compensation or the presence of spin canting\cite{Wang2024}. Our results establish that Mn incorporation effectively acts as a ``magnetic switch'' that suppresses this internal magnetic cancellation, thereby reconfiguring the exchange network and driving a transition toward a more robust ferromagnetic order.

\begin{figure}[htbp!]
    \centering
    \includegraphics[width=0.5\textwidth]{Fig.3.jpg}
    \vspace{0pt} 
    \caption{ Electrical transport properties of trigonal Cr$_5$Te$_8$ and Cr$_4$MnTe$_8$ single crystals. Temperature dependence of the in-plane longitudinal resistivity for (a) Cr$_5$Te$_8$ and (b) Cr$_4$MnTe$_8$, respectively. (c, d) Temperature dependence of the magnetoresistance (MR) under an applied field of 3~T for Cr$_5$Te$_8$ and Cr$_4$MnTe$_8$, respectively. The insets to (c) and (d) display the field-dependent MR curves measured at 5~K and 225~K, respectively.}
    \vspace{0pt} 
    \label{Fig. 3}
\end{figure}

Electrical transport measurements provide a sensitive probe into the intricate magnetic structure and scattering mechanisms of tr-Cr$_5$Te$_8$ and Cr$_4$MnTe$_8$. As illustrated in Figs.~3(a) and (b), both compounds exhibit characteristic metallic behavior across the measured temperature range\cite{Liu2018}. The $\rho(T)$ curves display pronounced anomalies, clearly resolved as peaks in the derivatives, which correspond to their respective magnetic ordering temperatures\cite{Lan2025}. The incorporation of Mn introduces enhanced impurity scattering, manifested as a significant elevation in the residual resistivity from 0.15~m$\Omega\cdot$cm to 0.50~m$\Omega\cdot$cm. Both crystals exhibit negative magnetoresistance (MR), a hallmark of itinerant ferromagnets in which an external magnetic field suppresses spin-disorder scattering\cite{Liu2022}, As evidenced in the insets of Figs.~3(c) and (d), the MR sweeps follow the Khosla--Fischer relation\cite{Khosla1970,Yuan2018}, confirming that spin fluctuations are the dominant source of scattering. Despite this shared origin, their temperature-dependent MR profiles reveal a significant divergence in spin dynamics. In tr-Cr$_5$Te$_8$ shown in Fig.~3(c), the MR magnitude decreases sharply upon cooling, from $-$3.6\% at 5~K to $-$1.1\% at 100~K. In stark contrast, Cr$_4$MnTe$_8$ shown in Fig.~3(d) maintains a nearly temperature-independent MR value of approximately $-$1.0\% up to 150~K. This suppression of low-temperature MR in the doped system suggests the realization of a more coherent spin alignment with diminished spin-disorder scattering, which is consistent with a transition from a frustrated magnetic state to a more robust ferromagnetic order\cite{Lan2025}. Finally, as both samples approach their respective magnetic ordering temperatures, they exhibit peak negative MR values of $-$4.7\% and $-$3.3\%, respectively, due to enhanced critical magnetic fluctuations\cite{Jiang2024}.

\begin{figure}[htbp!]
    \centering
    \includegraphics[width=0.45\textwidth]{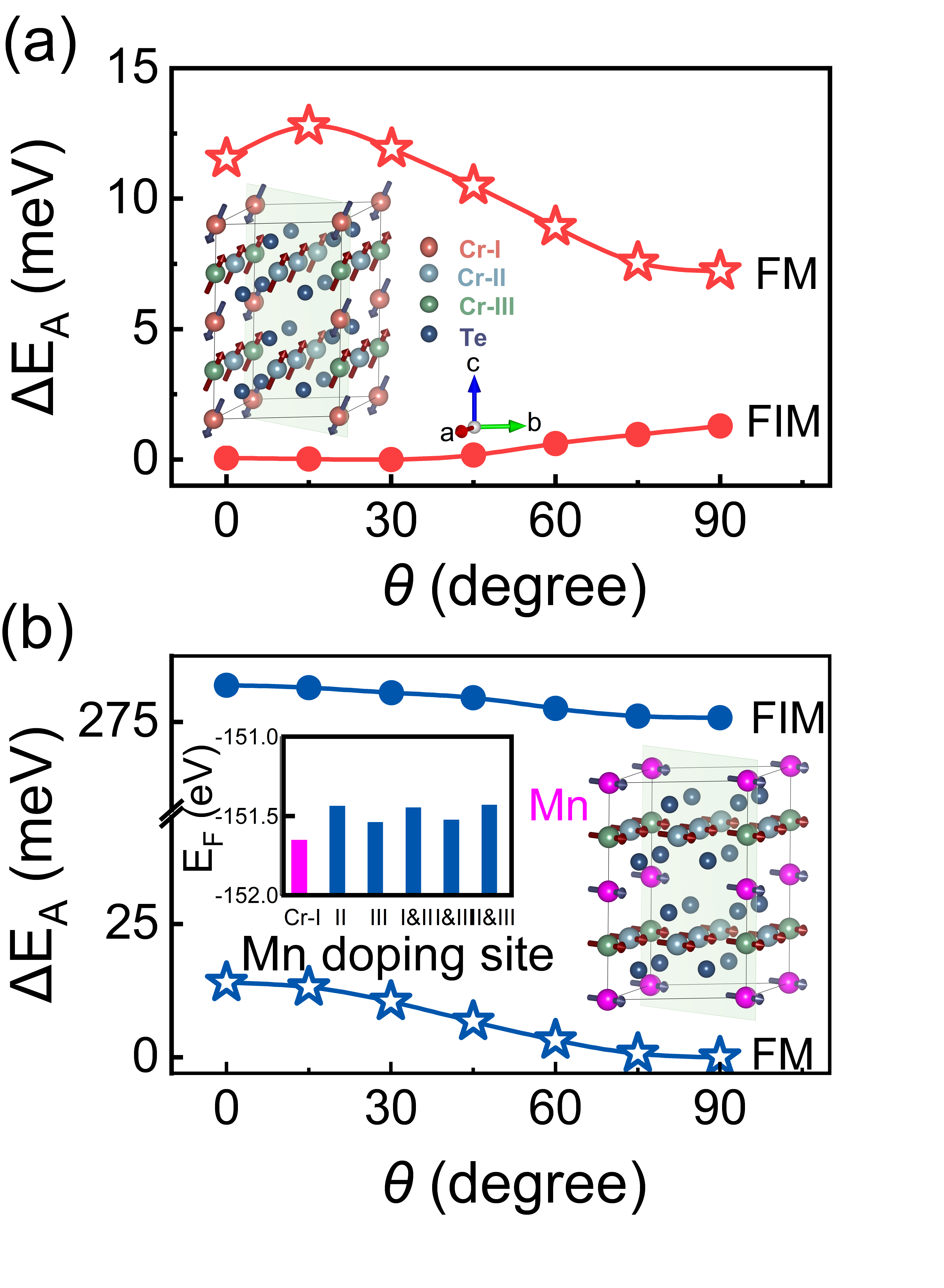}
    \vspace{0pt} 
    \caption{First-principles investigation of magnetic structures in pristine and Mn-doped $\text{Cr}_5\text{Te}_8$. (a) Magnetic anisotropy energy profiles for ferromagnetic and ferrimagnetic $\text{Cr}_5\text{Te}_8$ as a function of the spin orientation angle $\theta$ relative to the $c$-axis. The energy minimum occurs at $\theta = 30^\circ$, with the corresponding spin configuration illustrated in the inset. (b) Magnetic anisotropy energy profiles for ferromagnetic and ferrimagnetic $\text{Cr}_4\text{MnTe}_8$ versus $\theta$. The inset presents formation energies for Mn substitution at three inequivalent Cr sites in $\text{Cr}_5\text{Te}_8$, together with the calculated ground-state magnetic configuration of $\text{Cr}_4\text{MnTe}_8$.}
    \vspace{0pt} 
    \label{Fig. 4}
\end{figure}

Guided by our experimental observations, first-principles calculations were carried out to elucidate the magnetic ground states of trigonal Cr$_5$Te$_8$ and Cr$_4$MnTe$_8$. For Cr$_5$Te$_8$ in Fig.4(a), total-energy comparisons between ferromagnetic and ferrimagnetic configurations demonstrate that the ferrimagnetic state is consistently lower in energy, thus indicating it as the magnetic ground state\cite{Jiang2022,Purwar2024}. The magnetic anisotropy energy analysis further identifies the $c$-axis as the easy axis, marked by a shallow energy minimum with a 30$^\circ$ canting angle (energy penalty $<$ 0.1 meV/f.u.) and a subsequent steep energy ascent\cite{Purwar2024}. The calculated local magnetic moment of approximately $3.0~\mu_\text{B}/\text{Cr}$, when combined with the antiparallel spin alignment between fully and partially occupied Cr layers as well as the actual sample composition, yields a net average magnetic moment of approximately $1.98~\mu_\text{B}/\text{Cr}$. This result is in excellent agreement with our experimental value of 2.00 $\mu_\text{B}$/Cr in Fig. 2(c), providing definitive theoretical evidence for the intrinsic ferrimagnetic nature of the trigonal phase\cite{Tang2022}.

For Cr$_4$MnTe$_8$, the energetic preference for Mn substitution was first established. As shown in the inset of Fig.4(b), the configuration in which Mn atoms solely substitute the Cr-I sites within the vdW gaps exhibits the lowest formation energy. Based on this configuration, magnetic anisotropy energy calculations in Fig. 4(b) unambiguously reveal that the collinear ferromagnetic state becomes energetically more stable than any ferrimagnetic arrangement, indicating that Mn heterointercalation drives a phase transition from ferrimagnetic to ferromagnetic order. This observation aligns with the predicted strong ferromagnetic coupling between Cr and Mn sublattices in this class of transition metal tellurides\cite{Li2005,Li2011,Debnath2016,Zhang2018,Maniv2020}. The calculated local magnetic moments are approximately $3.5~\mu_\text{B}/\text{Mn}$ and $3.0~\mu_\text{B}/\text{Cr}$, respectively. Taking into account the actual composition of the sample, the average net magnetic moment is determined to be $2.94~\mu_\text{B}/\text{ion}$, which is in good agreement with the experimental saturation magnetization ($m_\text{S}$) value of $2.66~\mu_\text{B}/\text{ion}$ shown in  Fig. 2(d). This transition to a simplified collinear spin arrangement relieves the intrinsic spin compensation of the parent tr-Cr$_5$Te$_8$ lattice and rationalizes the suppressed low-temperature magnetoresistance observed in Fig.3(d). Nevertheless, the calculations predicting an in-plane ferromagnetic ground state conflict with the experimental determination of the $c$-axis as the easy axis in Fig. 2(c), a discrepancy that likely stems from the extreme sensitivity of magnetic anisotropy to subtle lattice strains.

Collectively, these findings demonstrate that heteroatomic dopants incorporated into the Cr sublattice function as atomic-scale ``magnetic switches'' capable of reconfiguring the spin exchange network without compromising structural integrity. This paradigm significantly advances the framework of intercalation-induced magnetism, establishing a rational design strategy for engineering high-temperature ferromagnetism in layered systems and facilitating their integration into emerging 2D spintronic technologies\cite{Maniv2020}.

\section{Conclusion}
In summary, we have performed a systematic investigation on the magnetic properties of tr-Cr$_5$Te$_8$ single crystals and their Mn-heterointercalated derivative, Cr$_4$MnTe$_8$. Mn intercalation effectively elevates the magnetic ordering temperature from 226 to 249 K and substantially enhances the saturation moment per magnetic ion from 2.00 to 2.66 $\mu_\text{B}$. Remarkably, the observed enhancement of the moment exceeds the nominal contribution expected from Mn ions alone. This result provides compelling evidence that Mn intercalation effectively eliminates the partial antiparallel alignment of the Cr spins intrinsic to the parent tr-Cr$_5$Te$_8$ lattice. By integrating magnetometry with first-principles calculations, we definitively establish tr-Cr$_5$Te$_8$ as a ferrimagnet with spins oriented approximately 30$^\circ$ relative to the $c$-axis. Furthermore, our calculations reveal that Mn ions preferentially occupy interstitial sites within the van der Waals gaps, functioning as a ``magnetic switch'' that drives a global transition from a ferrimagnetic to a ferromagnetic state. Collectively, these findings elucidate the magnetic structure of tr-Cr$_5$Te$_8$ and validate heterointercalation as a robust methodology for tailoring magnetic ordering in Cr$_x$Te$_y$ compounds, advancing their potential for integration into next-generation spintronic and topological quantum devices.

\begin{acknowledgments}
The authors gratefully acknowledge the financial support provided by the National Natural Science Foundation of China (Grant No.~51971087), the ``333 Talent Project'' of Hebei Province (Grant No.~C20231105), the Basic Research Project of Shijiazhuang Municipal Universities in Hebei Province (Grant No.~241790617A), the Central Guidance on Local Science and Technology Development Fund of Hebei Province (Grant No.~236Z7606G), and the Science Foundation of Hebei Normal University (Grant No.~L2024B08). These funding sources have been instrumental in facilitating the completion of this research.
\end{acknowledgments}

\end{document}